\documentclass[epjCONF]{svjour}
\usepackage{graphics}
\usepackage[varg]{txfonts} 
\usepackage[latin1]{inputenc}
\session-title{Conference Title, to be filled}
\begin{document}
\title{Magnetic fields in O-, B- and A-type stars on the main sequence}
\author{Maryline Briquet\inst{1}\fnmsep\inst{2}\fnmsep\thanks{\email{maryline.briquet@ulg.ac.be}}\fnmsep\thanks{F.R.S.-FNRS Postdoctoral Researcher, Belgium} }
\institute{Institut d'Astrophysique et de G\'eophysique, Universit\'e de Li\`ege, All\'ee du 6 Ao\^ut 17, B\^at B5c, 4000, Li\`ege, Belgium \and LESIA, Observatoire de Paris, CNRS UMR 8109, UPMC, Univ. Paris Diderot, 5 place Jules Janssen, 92195, Meudon Cedex, France}
 \abstract{In this review, the latest observational results on magnetic fields in main-sequence stars with radiative envelopes are summarised together with the theoretical works aimed at explaining them.} 
 \maketitle
 \section{Introduction}
Magnetic fields have first been discovered in the Sun \cite{hale08} and in the chemically peculiar A-type star 78~Virginis \cite{babcock47}. These two stars are representative of two major groups of magnetic stars, which are directly related to the dominant mechanism of heat transport in the outer layers of the stars. For the Sun and low-mass stars on the main sequence, the magnetic fields have complex surface structures and they are variable over time on various time scales. It is understood that those fields are generated by an ongoing solar-type dynamo, involving differential rotation and turbulence \cite{donati11}. For Ap stars, the fields are simple fields of globally dipole structure and they are extremely stable over time on long time scales. For these stars with radiative envelopes, it is proven that the fields are remnants of an early phase of the star-life and one speaks of fossil fields (see Sect.~\ref{fossil}). 

Over the past decade important progress has been achieved in the area of stellar magnetism. It is mainly due to the development of a new generation of high-performance spectropolarimeters, among which FORS\,1/2 at VLT (Paranal, Chile), ESPaDOnS at CFHT (Hawaii, USA), Narval at TBL (Pic du Midi, France), and HARPSpol at ESO 3.6-m (La Silla, Chile). At the present time, magnetic fields are also detected in pre-main sequence stars, hotter main-sequence stars, and more evolved stars. The detection of fields in these objects relies on the Zeeman effect, i.e. the splitting and polarisation of spectral lines produced by a magnetic field. By measuring the circular polarisation produced by the longitudinal component of the stellar field, it is now possible to detect fields down to a few Gauss in certain types of stars, for instance, in bright main-sequence A-type stars with sharp lines (see Sect.~\ref{Amag}).

Recent reviews on magnetic fields across the H-R diagram are given in \cite{mathys12} and \cite{landstreet13}. In what follows, an overview of our current observational and theoretical knowledge on magnetic main-sequence intermediate- and high-mass stars is presented.

\section{Observational results obtained with current spectropolarimetry}

\subsection{Magnetic field properties of intermediate- and high-mass stars}
\subsubsection{Two classes of magnetism among A-type stars}
\label{Amag}
It has been known for a long time that Ap/Bp stars are usually found to possess strong large-scaled organized magnetic fields. Current spectropolarimetric data give us new insights about main-sequence A-type stars. First, it is found that all Ap stars are magnetic \cite{auriere07}. The observed field is always higher than a limit of 100~G for the longitudinal field, which corresponds to a polar field of about 300~G. In addition, such strong fields are not observed in the A-type non Ap stars down to a few Gauss, which proves that the non-detection is not due to an instrumental limitation \cite{auriere10}. Finally, sub-Gauss fields are convincingly detected in the bright A-type star Vega and in a few other A-type stars as well \cite{lignieres09}, \cite{petit10}. Therefore, among intermediate-mass stars, there are two classes of magnetism: the Ap-like ``strong'' magnetism and the Vega-like ``ultra-weak'' magnetism, and between them, there is a ``magnetic desert'' \cite{lignieres14}.

\subsubsection{Diversity among O- and B-type stars}
For several years, intensive magnetic surveys of massive stars have been performed by different collaborations (e.g., Magori collaboration, \cite{hubrig11}, \cite{hubrig13}) and international consortia (e.g., MiMeS project -- Magnetism in Massive Stars \cite{wade14}, BOB collaboration -- B-fields in OB stars \cite{morel14}, BinaMIcs project -- Binarity and Magnetic Interactions in various classes of Stars \cite{alecian14}). These efforts provide us with the first observational view of the magnetic properties of hot stars. Both MiMeS and BOB surveys conclude to an incidence of magnetic detection in massive stars similar to that of Ap stars. A fraction between 5 to 10 percent of main-sequence stars with radiative envelopes hosts detectable fields, regardless of the spectral type. The topologies of magnetic massive stars are also usually similar to those of Ap stars. They are oblique dipole very stable over years but complex structures are also observed \cite{donati06}, \cite{kochukhov10}. 
Among magnetic massive stars, there are very diverse objects: slowly rotating and rapidly rotating stars, strong fields and weaker fields, stars with surface chemical peculiarity, and others with normal chemical abundances. Magnetics fields are detected in non-pulsating stars and also in pulsating stars \cite{neiner03}, \cite{hubrig06}. Moreover, there is no correlation between the observed stellar properties. In particular, no correlation is found between the stellar rotation rate and the stellar field strength, contrary to what is observed in main-sequence stars with convective envelopes. 

\subsection{Weak magnetic fields among O and B-type stars}

Among O and B-type stars, there is the extension of the Ap stars, i.e. stars with a typical polar strength of the order of kiloGauss (``strong'' fields). In addition, evidence for fields with polar field of the order of hundred Gauss (``weak'' fields) is increasingly being observed. So far, such weak fields seem to be observed in HD~37742 ($\zeta$ OriAa) \cite{blazere14}, $\tau$~Sco \cite{donati06}, $\epsilon$~CMa \cite{fossati14}, $\beta$~CMa \cite{fossati14}, and $\zeta$~Cas \cite{briquet}. The case of the pulsating $\beta$~Cep star $\beta$~CMa illustrates that magnetic fields in massive stars might be more ubiquitous than currently known. The first magnetic analyses by independent teams using two different instruments and different methods for data analysis led to a non-detection in this star \cite{hubrig09}, \cite{silvester09}. However, the typical signature in Stokes~V of a weak ($<$30~G in absolute value) longitudinal magnetic field has recently been revealed by HARPSpol data \cite{fossati14}. The polar field strength of this star is found to be about 100~G, which is much weaker than what is usually determined for other massive stars, and also lower than what is found in the typical Ap-type stars. This indicates a possible lack of a ``magnetic desert'' in massive stars, contrary to what is observed in main-sequence A-type stars.

\subsection{Asteroseismology of magnetic massive stars}

A way to investigate the effects of magnetic fields on stellar structure and evolution is to perform asteroseismic studies of magnetic objects. The first asteroseismic modelling of a magnetic massive star was performed on the B-type pulsator $\beta$~Cephei \cite{shibahashi00}. This work dates back to more than a decade ago and was the only one available until recently. With the aim to perform an asteroseismic modelling of the magnetic $\beta$~Cep star V2052~Oph, intensive multisite photometric and spectroscopic campaigns were set up \cite{handler12}, \cite{briquet12}. The outcome of the modelling shows that the magnetic field observed in the star inhibits mixing in its radiative zone. The field strength observed in this star \cite{neiner12} is 6 to 10 higher than the critical field limit needed to inhibit mixing as determined from theory \cite{mathis05}, \cite{zahn11}. 

Thanks to photometric data assembled by the CoRoT satellite complemented by ground-based spectroscopy and spectropolarimetry, HD~43317 was discovered to be a single magnetic B-type hybrid pulsator with a wealth of observational information that can be used to calibrate models of hot magnetic stars \cite{papics12}, \cite{briquet13}. As detailed in \cite{neiner14}, current opportunities to perform such studies are being provided by the BRITE and K2 satellites.

\section{Nature and origin of magnetic fields in stars with radiative envelopes}

\subsection{Intermediate- and high-mass stars on the main sequence}

\subsubsection{Core dynamo fields and dynamo process in radiative zones}
First, it is presently believed that the observed surface magnetic fields cannot be core dynamo fields. Numerical simulations show that dynamo activity in the core is taking place but the time needed for the field to be visible at the surface would be longer than the lifetime of the star \cite{charbonneau01}, \cite{brun05}. Moreover, such fields do not have the observed topology. Second, it has been proposed that a dynamo process could operate in radiative envelopes and a toroidal field could be generated by differential rotation \cite{braithwaite06}. However, another study concluded that it is not possible to excite/maintain a Tayler-Spruit dynamo in a radiative envelope \cite{zahn07}. If such a dynamo process in radiative envelopes would exist, it would imply a correlation between stellar rotation and field strength that is not observed. For this reason, it can be concluded that such fields are not those observed at the surface of massive stars.

\subsubsection{Fields produced by sub-surface convective layers}
Massive stars have thin sub-surface convective layers that are induced by the iron opacity bump near the surface. Therefore, it is plausible that a dynamo process can operate in the sub-surface convection zones. Moreover, these fields could reach the stellar surface and it is shown that increasing strengths are found for increasing masses and towards the end of the main sequence \cite{cantiello11}. The potential effects of these fields would be larger for the hottest stars. Such fields would produce small-scaled field structure and when emerging at the surface would produce hot bright spots at the surface of massive stars. So far there are no direct observations of these fields because they lead to small-scaled structure of weak strengths. However, there is increasing evidence of indirect observations of such fields in hot stars thanks to space-based photometry. CoRoT and Kepler light curves have revealed rotational modulation in several hot OB stars that might be due to such bright spots \cite{papics11}, \cite{balona11}. Recently, a convincing case of co-rotating bright spots on an O-type star has also been reported thanks to data taken by the MOST satellite \cite{ramiaramanantsoa14}. Again, the current opportunities to study this kind of magnetic activity in massive stars are the K2 and BRITE satellites. Such studies are potentially important as these fields induced by sub-surface convective layers are very likely to exist and they may directly affect the evolution of the more massive stars by altering the stellar wind mass-loss and enhancing the loss of angular momentum. 

\subsubsection{Fossil fields}
\label{fossil}
An explanation of the observed fields of Ap stars already proposed a long time ago is the fossil origin. The fossil origin suggests that magnetic fields reside inside the star without being continuously renewed. Therefore, these fields have been formed during an early phase of the life of the star. The strong observational argument supporting this view is that the observed large-scaled organised fields are extremely stable over time. This hypothesis was very plausible but difficult to prove. Early analytical works could show that certain field configurations are unstable, for instance all axisymmetric fields which are either purely poloidal or purely toroidal are unstable. Then, it was postulated that a configuration with both a torodial component and a poloidal component (a mixed configuration) would be stable. In recent years, semi-analytical works and numerical simulations have definitely proven that mixed fields can be stable inside radiative envelopes \cite{braithwaitenordlund06}, \cite{duez10}. Numerical simulations show that an initial arbitrary field in a star evolves towards a stable mixed configuration over relatively short time scales and afterwards it continues to evolve on very long timescales. Moreover, these stable axisymmetric configurations can reproduce those observed at the stellar surface where they are seen as dipole. Other simulations show that non-axisymmetric configurations can also be stable and could explain the more complex structure observed in some massive stars \cite{braithwaite08}. 

\subsubsection{Vega-like magnetism}
Several scenarios to explain the Vega-like magnetism (sub-Gauss fields in main-sequence A-type stars) have been proposed but more observational and numerical works are needed in order to discriminate between the different suggestions. A first possibility is that these fields are fossil fields but, because of the weakness of the field, the time needed to reach a dipolar equilibrium is longer than the current age of the star. Therefore, we would observe fields that are still evolving towards an equilibrium state. Such fields have been called ``failed fossils'' \cite{braithwaite13}. Another suggestion is that the field is too weak to freeze possible differential rotation. As a consequence a strong toroidal field is produced and the configuration becomes unstable. In this scenario, the lower limit for Ap-like magnetism would correspond to the critical dipolar field which separates the stable from the unstable configurations \cite{lignieres14}. A third explanation would be that sub-surface convective layers induced by the helium opacity bump could generate a magnetic field.
 
\subsection{Herbig pre-main sequence stars}
An observational way to test whether magnetic fields on the main sequence are generated during an early phase of the star-life is to investigate the incidence and properties of pre-main sequence stars. With the aim to test the fossil origin of magnetic intermediate-mass main-sequence stars, surveys dedicated to the study of the magnetic properties of Herbig stars have been performed over the last years. About 6 percent of the Herbig stars are magnetic. The topologies are simple and stable over years. The properties of the magnetic fields of the studied Herbig stars, which are fully radiative stars, indicate that the fields observed on the main sequence are already present during the Herbig phase \cite{alecian13}. Consequently, the fields must have been shaped before the Herbig phase, at an earlier stage of evolution. 

In conclusion, the fossil field hypothesis for intermediate- and high-mass stars on the main sequence has been proven both theoretically and observationally. What remains speculative and debated is the exact way in which such fields are formed. Several scenarios have been proposed but further observational efforts are needed to test them. The magnetic field could be the capture of the magnetic field of the interstellar medium from which the star contracts \cite{moss03}, or it could be generated by dynamo action during a pre-main sequence stage during which convection is active \cite{arlt11}, or it could also be the result of the merging of components of a close binary \cite{ferrario09}, \cite{tutukov10}.
%

%
%


\begin{thebibliography}{}
\bibitem{hale08}Hale G.E., ApJ \textbf{28}, (1908) 315
\bibitem{babcock47}Babcock H.W., ApJ \textbf{105}, (1947) 105
\bibitem{donati11}Donati J.-F., IAUS \textbf{271}, (2011) 23
\bibitem{mathys12}Mathys G., ASPC \textbf{462}, (2012) 295
\bibitem{landstreet13}Landstreet J.D., EAS \textbf{63}, (2013) 67
\bibitem{auriere07}Auri\`ere M., Wade G.A., Silvester J., et al., A\&A \textbf{475}, (2007) 1053
\bibitem{auriere10}Auri\`ere M., Wade G.A., Ligni\`eres F., et al., A\&A \textbf{523}, (2010) 40
\bibitem{lignieres09}Ligni\`eres F., Petit P., B\"ohm T., Auri\`ere M., A\&A \textbf{500}, (2009) L41
\bibitem{petit10}Petit P., Ligni\`eres F., Wade G.A., et al., A\&A \textbf{523}, (2010) 41
\bibitem{lignieres14}Ligni\`eres F., Petit P., Aurière M., Wade G.A., B\"ohm T., IAUS \textbf{302}, (2014) 338
\bibitem{hubrig11}Hubrig S., Sch\"oller M., Kharchenko N.V., A\&A \textbf{528}, (2011) 151
\bibitem{hubrig13}Hubrig S., Sch\"oller M., Ilyin I., et al., A\&A \textbf{551}, (2013) 33 
\bibitem{wade14}Wade G.A., Grunhut J., Alecian E., et al., IAUS \textbf{302}, (2014) 265
\bibitem{morel14}Morel T., Castro N., Fossati L., et al., Msngr \textbf{157}, (2014) 27
\bibitem{alecian14}Alecian E., Neiner C., Wade G.A., et al., IAUS \textbf{307}, in press (arXiv:1409.1094)
\bibitem{donati06}Donati J.-F., Howarth I.D., Jardine M.M., et al., MNRAS \textbf{370}, (2006) 629
\bibitem{kochukhov10}Kochukhov O. \& Wade G.A., A\&A \textbf{513}, (2010), 13
\bibitem{neiner03}Neiner C., Henrichs H.F., Floquet M., et al., A\&A \textbf{411}, (2003) 565
\bibitem{hubrig06}Hubrig S., Briquet M., Sch\"oller, et al., MNRAS \textbf{369}, (2006) 61
\bibitem{blazere14}Blaz\`ere A., Neiner C., Bouret J-C., Tkachenko A., IAUS \textbf{307}, in press (arXiv:1408.0178)
\bibitem{fossati14}Fossati L., Castro N., Morel T., et al., A\&A, in press (arXiv:1411.6490)
\bibitem{briquet}Briquet M., Neiner C., Petit P., Leroy B., de Batz B., in prep.
\bibitem{hubrig09}Hubrig S., Briquet M., De Cat P. et al., AN \textbf{330}, (2009) 317
\bibitem{silvester09}Silvester J., Neiner C., Henrichs H.F., et al., MNRAS \textbf{398}, (2009) 1505
\bibitem{shibahashi00}Shibahashi H. \& Aerts C., ApJ \textbf{531}, (2000) 143
\bibitem{handler12}Handler G., Shobbrook R.R., Uytterhoeven K., et al., MNRAS \textbf{424}, (2012) 2380
\bibitem{briquet12}Briquet M., Neiner C., Aerts C., et al., MNRAS \textbf{427}, (2012) 483
\bibitem{neiner12}Neiner C., Alecian E., Briquet M., et al., A\&A \textbf{537}, (2012) 148 
\bibitem{mathis05}Mathis S. \& Zahn. J.-P. A\&A \textbf{440}, (2005) 653
\bibitem{zahn11}Zahn J.-P., IAUS \textbf{272}, (2011) 14
\bibitem{papics12}P\'apics P.I., Briquet M., Baglin A., et al., A\&A \textbf{542}, (2012) 55
\bibitem{briquet13}Briquet M., Neiner C., Leroy B., P\'apics P.I., A\&A \textbf{557}, (2013) 16
\bibitem{neiner14}Neiner C., Briquet M., Mathis S., Degroote P., IAUS \textbf{307}, in press (arXiv:1407.8087)
\bibitem{charbonneau01}Charbonneau P. \& MacGregor K.B., ApJ \textbf{559}, (2001) 1094
\bibitem{brun05}Brun A.S., Browning M.K., Toomre J., ApJ \textbf{629}, (2005) 461	
\bibitem{braithwaite06}Braithwaite J., A\&A \textbf{449}, (2006) 451
\bibitem{zahn07}Zahn J.-P., Brun A.S., Mathis S., A\&A \textbf{474}, (2007) 145
\bibitem{cantiello11}Cantiello M. \& Braithwaite J., A\&A \textbf{534}, (2011) 140
\bibitem{papics11}P\'apics P.I., Briquet M., Auvergne M., et al., A\&A \textbf{528}, (2011) 123
\bibitem{balona11}Balona L.A., Pigulski A., De Cat P., et al., MNRAS \textbf{413}, (2011) 2403 
\bibitem{ramiaramanantsoa14}Ramiaramanantsoa T., Moffat A.F.J., Chen\'e, A.-N., et al., MNRAS \textbf{441}, (2014) 910 
\bibitem{braithwaitenordlund06}Braithwaite J. \& Nordlund A., A\&A \textbf{450}, (2006) 1077
\bibitem{duez10}Duez V. \& Mathis S., A\&A \textbf{517}, (2010) 58
\bibitem{braithwaite08}Braithwaite J., MNRAS \textbf{386}, (2008) 1947
\bibitem{braithwaite13}Braithwaite J. \& Cantiello M., MNRAS \textbf{428}, (2013) 2789
\bibitem{alecian13}Alecian E., Wade G.A., Catala C., et al., MNRAS \textbf{429}, (2013) 1001
\bibitem{moss03}Moss D., A\&A \textbf{403}, (2003) 693
\bibitem{arlt11}Arlt R. \& R\"udiger G., MNRAS \textbf{412}, (2011) 107
\bibitem{ferrario09}Ferrario L., Pringle J.E., Tout C.A., Wickramasinghe D.T., MNRAS \textbf{400}, (2009) L71
\bibitem{tutukov10}Tutukov A.V. \& Fedorova A.V., ARep \textbf{54}, (2010) 156

\end{thebibliography}
\end{document}